\documentclass[aps,prc,twocolumn,showpacs,superscriptaddress]{revtex4-1}
\usepackage{graphicx}
\usepackage{amsmath,amssymb}
\usepackage{hyperref}
\usepackage{color}
\usepackage{multirow}
\usepackage{ulem}

\newcommand{\sbar}[1]{\ooalign{\hfil/\hfil\crcr$#1$}}

\begin{document}

\title{Nuclear structure in Parity Doublet Model}

\author{Myeong-Hwan~Mun}
\affiliation{Department of Physics and Origin of Matter and Evolution of Galaxy Institute, Soongsil University, Seoul 06978,  Korea}

\author{Ik~Jae~Shin}
\affiliation{Rare Isotope Science Project, Institute for Basic Science, Daejeon 34000, Korea}

\author{Won-Gi~Paeng}
\affiliation{Rare Isotope Science Project, Institute for Basic Science, Daejeon 34000, Korea}

\author{Masayasu Harada}
\affiliation{Department of Physics, Nagoya University, Nagoya, 464-8602, Japan}
\affiliation{ Kobayashi-Maskawa Institute for the Origin of Particles and the Universe, Nagoya University, Nagoya, 464-8602, Japan}
\affiliation{Advanced Science Research Center, Japan Atomic Energy Agency, Tokai 319-1195, Japan}

\author{Youngman~Kim}
\affiliation{Center for Exotic Nuclear Studies, Institute for Basic Science, Daejeon 34126, Korea}

\date{\today}

\begin{abstract}
Using an extended parity doublet model with the hidden local symmetry, we study some properties of nuclei in the mean field approximation to see
 if the parity doublet model could reproduce nuclear properties and also to estimate the value of the chiral invariant nucleon mass $m_0$ preferred by nuclear structure.
 We first determine our model parameters using the inputs from free space and from nuclear matter properties. Then, we study some basic nuclear properties
 such as the nuclear binding energy with  several different choices of the chiral invariant mass.
We observe that our results approach the experimental values as $m_0$ is increased until $m_0=700$~MeV
and start to deviate more from the experiments afterwards with $m_0$ larger than $m_0=700$~MeV.
From this observation, we conclude that $m_0=700$~MeV is preferred by nuclear properties.
We then calculate some properties of several selected nuclei with $m_0=700$ MeV and compare them with experiments.
Finally, we study the neutron-proton mass difference in some nuclei.
\end{abstract}


\maketitle

\section{Introduction}
Nuclei are interesting and important quantum finite many-body systems,  providing a solid testing ground  for our understanding of the strong interactions  and
many-body techniques. In principle, we should be able to understand nuclei in terms of quarks and gluons in the framework of quantum chromodynamics (QCD).
For instance, it will be especially interesting to investigate how QCD vacuum properties  are reflected in the properties of
nuclear matter and finite nuclei. This is, however, a formidable task to achieve  due to the non-perturbative nature of the strong interaction at low energies.
Since the nucleus is commonly believed to be composed of the protons and neutrons and it is highly nontrivial to understand nuclei in terms of quarks and gluons in the framework of QCD, it is natural to view the nucleus as a collection of interacting protons and neutrons. In addition, in the light of effective theory, thanks to the QCD separation scale due to confinement or spontaneous chiral symmetry breaking, using protons and neutrons as relevant degrees of freedom for nuclei would be still desirable.

Studying the origin of hadron masses is one of important problems in nuclear physics. As it is well known,
the current quark mass could explain roughly 2\% of the nucleon mass. A picture in nuclear physics
is that the nucleon mass in the chiral limit could be explained by quark-antiquark condensates in QCD vacuum, i.e., spontaneous chiral symmetry breaking.
In the parity doublet model~\cite{Detar:1988kn}, however, the nucleon mass has a piece called the chiral invariant mass $m_0$ that may have something to do with
QCD trace anomaly. In Ref.~\cite{Detar:1988kn}, from the decay of $N^\ast$, $N^\ast (1535)\rightarrow N+\pi$, the value of the chiral invariant mass $m_0$
was determined as $m_0=270$~MeV.

Dense matter, which is intimately related to heavy ion collisions, nuclear structure and neutron stars, has been extensively investigated in the parity doublet models~\cite{Hatsuda:1988mv, Zschiesche:2006zj, Dexheimer:2007tn,Sasaki:2010bp,Gallas:2011qp,Sasaki:2011ff, Steinheimer:2011ea,Benic:2015pia, Motohiro:2015taa},
while nuclear structure study in parity doublet models has not been made.
In Ref.~\cite{Zschiesche:2006zj}, the chiral invariant mass was estimated as $m_0\sim 800$~MeV from nuclear matter properties, especially incompressibility.
In Ref.~\cite{Gallas:2011qp}, the chiral invariant mass $m_0$ was re-expressed as the sum of the contributions from tetraquark and gluon condensates.
An extended parity doublet model~\cite{Motohiro:2015taa} reasonably reproduces the properties of nuclear matter with the chiral invariant nucleon mass in the range from $500$ to $900$~MeV.
As discussed in Ref.~\cite{Harada:2019oaq}, it is expected that a larger $m_0$ implies a smaller Yukawa coupling of the $\sigma$ field to nucleons, since a part of the nucleon mass from the chiral symmetry breaking is smaller.
The attractive force by the $\sigma$ should be balanced by the repulsive force mediated by $\omega$ at the saturation density, so that the $\omega$ contribution is smaller for large $m_0$.
In the higher density region, $\omega$ contribution is expected to grow while $\sigma$ contribution decreases.
As a result of the repulsive nature of the $\omega$, the equation of state is softer for large $m_0$~\cite{Harada:2019oaq}.

A promising microscopic theoretical tool for nuclear matter and (medium-mass and heavy) nuclei is energy density functional (EDF) theory~\cite{Bender:2003jk}.
With a few parameters, density functional theory  provides a successful description of ground-state properties of
spherical and deformed nuclei.
Self-consistent relativistic mean field theory is a useful method to obtain the covariant energy density functionals.
The covariant EDF includes the nucleon spin degree of freedom naturally  and, therefore, can consistently explain the nuclear spin-orbit potential.
For the details of covariant EDF, we refer to Refs.~\cite{Vretenar:2005zz, Meng:2005jv, Meng:2016, Schunck:2019}.

In this work, using the parity doublet model developed in Ref.~\cite{Motohiro:2015taa}, we study the properties of nuclei in self-consistent relativistic mean field theory
to see if a parity doublet model works for nuclear properties and to find out
the value of the chiral invariant mass preferred by nuclear structures.
As a first attempt, we will focus on the properties of stable nuclei in the present study.

The extended parity doublet model~\cite{Motohiro:2015taa}, where the chiral invariant nucleon mass is in the range from $500$ to $900$~MeV, is briefly described in Section~\ref{MKH model} and the results are given in Section~\ref{results}. We then summarize the present work in Section~\ref{summary}.

\section{Parity doublet model with hidden local symmetry} \label{MKH model}
In this section we study finite nuclei in the context of the parity doublet model~\cite{Detar:1988kn,Jido:2001nt,Gallas:2009qp,Paeng:2011hy}.
To investigate nuclear matter or finite nuclei using a parity doublet model, either isospin symmetric or asymmetric, one needs to introduce vector mesons in the model.
A convenient way to do that, which respects chiral symmetry, is to use the Hidden Local Symmetry (HLS)~\cite{Bando:1987br,Harada:2003jx}.
In Ref.~\cite{Motohiro:2015taa}, a parity doublet model with the HLS was constructed for an asymmetric nuclear matter study.
It was shown in Ref.~\cite{Motohiro:2015taa} that the phase structure of cold dense matter depends on the value of the chiral invariant mass and also on
 isospin asymmetry.
In this section, we use the extended parity doublet model constructed in Ref.~\cite{Motohiro:2015taa}.

We start with the Lagrangian,
\begin{eqnarray}
{\cal L} =&&~\bar{\psi}_1 i \sbar{\partial} \psi_1 + \bar{\psi}_2 i \sbar{\partial}\psi_2 + m_0 (\bar{\psi}_2 \gamma_5 \psi_1 - \bar{\psi}_1 \gamma_5 \psi_2) \nonumber\\
&& + g_1\bar{\psi}_1 (\sigma + i \gamma_5 \vec{\tau} \cdot \vec{\pi})\psi_1 + g_2\bar{\psi}_2(\sigma - i \gamma_5 \vec{\tau} \cdot \vec{\pi})\psi_2 \nonumber\\
&& -g_{\omega N\!N} \bar{\psi}_1 \gamma_\mu \omega^\mu \psi_1 -g_{\omega N\!N} \bar{\psi}_2 \gamma_\mu \omega^\mu \psi_2 \nonumber\\
&& -g_{\rho N\!N} \bar{\psi}_1 \gamma_\mu \vec{\rho}^{\,\mu} \cdot \vec{\tau} \psi_1 -g_{\rho N\!N} \bar{\psi}_2 \gamma_\mu \vec{\rho}^{\,\mu} \cdot \vec{\tau} \psi_2 \nonumber\\
&& -e \bar{\psi}_1 \gamma^\mu A_\mu \frac{1-\tau_3}{2} \psi_1 -e \bar{\psi}_2 \gamma^\mu A_\mu \frac{1-\tau_3}{2} \psi_2 + {\cal L}_M\, ,~~
\end{eqnarray}
where the baryon fields $\psi_1$ and $\psi_2$ transform  as
\begin{eqnarray}
&&\psi_{1R} \rightarrow R\psi_{1R}\, , \quad \psi_{1L} \rightarrow L\psi_{1L}\, , \nonumber \\
&&\psi_{2R} \rightarrow L\psi_{2R}\, , \quad \psi_{2L} \rightarrow R\psi_{2L}\, ,
\end{eqnarray}
with $L$ and $R$ being the elements of  $SU(2)_L$ and  $SU(2)_R$ chiral symmetry group, respectively.
The meson part of Lagrangian is given by
\begin{eqnarray}
{\cal L}_M =&&~\frac{1}{2} \partial_\mu \sigma \partial^\mu \sigma + \frac{1}{2} \partial_\mu \vec{\pi} \cdot \partial^\mu \vec{\pi} \nonumber\\
&& - \frac{1}{4} \Omega_{\mu\nu} \Omega^{\mu\nu} -\frac{1}{4} \vec{R}_{\mu\nu} \cdot \vec{R}^{\mu\nu} -\frac{1}{4} F_{\mu\nu} F^{\mu\nu} \nonumber \\
&& + \frac{\bar{\mu}^2}{2} (\sigma^2 + \vec{\pi}^2) -\!\frac{\lambda}{4}(\sigma^2 + \vec{\pi}^2)^2 +\!\frac{\lambda_6}{6}(\sigma^2 + \vec{\pi}^2)^3 + \epsilon\sigma \nonumber\\
&& + \frac{1}{2}m_\omega^2 \omega_\mu \omega^\mu  + \frac{1}{2}m_\rho^2 \vec{\rho}_\mu \cdot \vec{\rho}^{\,\mu}
\end{eqnarray} with
\begin{eqnarray}
\Omega_{\mu\nu}&=&\partial_\mu \omega_\nu -\partial_\nu \omega_\mu\, , \nonumber\\
\vec{R}_{\mu\nu}&=&\partial_\mu \vec{\rho}_\nu -\partial_\nu \vec{\rho}_\mu\, , \nonumber\\
F_{\mu\nu}&=&\partial_\mu A_\nu -\partial_\nu A_\mu\,.
\end{eqnarray}
In this work, we adopt a mean field approximation and consider the following mean fields for mesons: $\sigma\rightarrow \langle \sigma \rangle$,
$\omega_\mu\rightarrow \delta_{\mu 0}\langle \omega_0 \rangle$, and $\rho_{i \mu}\rightarrow \delta_{i3}\delta_{\mu 0}\langle \rho_0^3 \rangle$.
The pion mass $m_\pi$, $\sigma$ meson mass $m_\sigma$, and pion decay constant $f_\pi$ are given by
\begin{eqnarray}
m_\pi^2&=&\lambda\langle\sigma\rangle^2-\bar\mu^2-\lambda_6\langle\sigma\rangle^4\, ,\nonumber \\
m_\sigma^2&=&3\lambda\langle\sigma\rangle^2-\bar\mu^2-5\lambda_6\langle\sigma\rangle^4\, ,\nonumber \\
f_\pi&=&\langle\sigma\rangle\, .
\end{eqnarray}
After diagonalization of the baryon mass terms, we get the mass of the nucleon field $N$, which corresponds to one of
the mass eigenstates, as
\begin{equation}
m_N = \frac{1}{2} \Big(\sqrt{(g_1+g_2)^2 \langle \sigma \rangle^2 +4m_0^2} - (g_1-g_2)\langle \sigma \rangle \Big)\, .  \label{mass}
\end{equation}

The equations of motion (EoM) for the stationary mean fields $\tilde{\sigma}$, $\omega_0$, $\rho_0^3$ and $A_0$ read
\begin{eqnarray}
\Big(-\vec{\nabla}^2 + m_\sigma^2\Big)\langle \tilde{\sigma}(\vec{x}) \rangle &=& - \bar{N}(\vec{x})N(\vec{x}) \left.\frac{\partial\, m_N( \tilde{\sigma})}{\partial \tilde{\sigma} }\right|_{\tilde{\sigma}=\langle \tilde{\sigma}(\vec{x}) \rangle}~\nonumber\\
&&+\left(-3 f_\pi \lambda+10 f_\pi^3 \lambda_6\right)\langle \tilde{\sigma}(\vec{x}) \rangle^2 \nonumber\\
&&+\left(-\lambda+10 f_\pi^2\lambda_6\right)\langle \tilde{\sigma}(\vec{x}) \rangle^3 \nonumber\\
&&+5 f_\pi\lambda_6\langle \tilde{\sigma}(\vec{x}) \rangle^4+\lambda_6\langle \tilde{\sigma}(\vec{x}) \rangle^5\, ,\label{EoMsig} \\
\Big(-\vec{\nabla}^2 + m_\omega^2 \Big)\langle \omega_0(\vec{x}) \rangle &=& g_{\omega N\!N}N^\dagger(\vec{x}) N(\vec{x})\,, \label{EoMome}\\
\Big(-\vec{\nabla}^2 + m_\rho^2 \Big)\langle \rho_0^3(\vec{x}) \rangle &=& g_{\rho N\!N}N^\dagger(\vec{x})\tau^3 N(\vec{x})\,, \label{EoMrho}\\
-\vec{\nabla}^2 \langle A_0(\vec{x}) \rangle &=& e N^\dagger(\vec{x}) \frac{1-\tau_3}{2} N(\vec{x})\,. \label{EoMphoton}
\end{eqnarray}
Note here that for calculational handiness we take the shift $\sigma = f_\pi + \tilde{\sigma}$.
Since we are interested in finite nuclei, we will not consider the EoM for the parity partner of the nucleon, $N^\ast(1535)$, which does not form its Fermi sea
near the saturation density. In addition, since our primary goal here is to see if the parity doublet model can explain some basic nuclear properties such as the binding energy, we will not consider pairing correlations which are essential for odd-even staggering in nuclear properties. For instance, according to the semi-empirical mass formula, the contribution from the pairing term to the binding energy per nucleon of $^{58}$Ni is only about $0.03$~MeV.

The EoM for the nucleon is given by
\begin{equation}
\big[ \vec{\alpha} \cdot \vec{p} + \beta\, m_N(\langle \tilde{\sigma}(\vec{x}) \rangle)  +V(\vec{x}) \big]N_i(\vec{x}) = \epsilon_i N_i(\vec{x})\,, \label{nucleon}
\end{equation}
where $N_i$ is the single-particle wave function and
\begin{equation}
V(\vec{x})=g_{\omega N\!N} \langle\omega_0(\vec{x})\rangle + g_{\rho N\!N}\langle\rho_0^3(\vec{x})\rangle\tau^3+e \frac{(1-\tau_3 )}{2} \langle A_0(\vec{x})\rangle\,.
\end{equation}
With assuming the spherical shape of the nucleus, we can solve the Eqs.~(\ref{EoMsig})-(\ref{EoMphoton}) and Eq.~(\ref{nucleon})   simultaneously to obtain the energy
\begin{equation}
E = \int d^3x \, {\cal H}(\vec{x}) \,.
\end{equation}
After subtracting out the vacuum contribution,  we write the Hamiltonian density ${\cal H}(\vec{x})$  in the mean field approximation as
\begin{widetext}
\begin{eqnarray}
{\cal H} &=& \bar{N} \left( -i\gamma^i \partial_i + m_N \right) N + g_{\omega N\!N} \langle \omega_0 \rangle N^\dagger N + g_{\rho N\!N} \langle \rho_0^3 \rangle N^\dagger \tau^3 N + e \langle A_0 \rangle N^\dagger \frac{1-\tau_3}{2}N \nonumber \\
&&-\frac{1}{2} \partial^i\langle \tilde{\sigma} \rangle \partial_i \langle \tilde{\sigma} \rangle +\frac{1}{2} \partial^i \langle \omega_0 \rangle \partial_i \langle \omega_0 \rangle +\frac{1}{2} \partial^i \langle \rho_0^3 \rangle \partial_i \langle \rho_0^3 \rangle +\frac{1}{2} \partial^i \langle A_0 \rangle \partial_i \langle A_0 \rangle \nonumber\\
&& -\frac{\bar{\mu}^2}{2} \left[ \left( f_\pi +\langle \tilde{\sigma} \rangle \right)^2 -f_\pi^2 \right] + \frac{\lambda}{4} \left[ \left( f_\pi +\langle \tilde{\sigma} \rangle \right)^4 -f_\pi^4 \right] - \frac{\lambda_6}{6} \left[ \left( f_\pi +\langle \tilde{\sigma} \rangle \right)^6 -f_\pi^6 \right] -\epsilon \langle \tilde{\sigma} \rangle \nonumber\\
&&-\frac{1}{2}m_\omega^2 \langle \omega_0 \rangle^2 -\frac{1}{2}m_\rho^2 \langle \rho_0^3 \rangle^2\,. \label{Hamil}
\end{eqnarray}
\end{widetext}
Then, the binding energy (BE) per nucleon  is given by
\begin{equation}
{\rm BE}/{A} = -\frac{E}{A} +m_N\,.
\end{equation}

\section{Results}\label{results}
Following Ref.~\cite{Motohiro:2015taa}, we determine the free parameters in our model using the inputs in free space listed in Table~\ref{table0} and nuclear matter properties.
\begin{table}[h]
\caption{The inputs from free space (in MeV).}
\begin{tabular}{c c c c c c}
\hline
\hline
~~$m_N$ &~~~$m_{N^\ast}$  &~~~$m_\omega$ &~~~$m_\rho$ &~~~$f_\pi$ &~~~$m_\pi$~~\\
\hline
~~$939$ &~~~$1535$ &~~~$783$      &~~~$776$    &~~~$93$    &~~~$138$~~ \\
\hline
\hline
\end{tabular}
\label{table0}
\end{table}
The nuclear matter properties used to fix the parameters are given by
\begin{eqnarray}
&&\frac{E}{A} - m_N = -16~{\rm MeV}\,, \quad n_0 = 0.16~{\rm fm}^{-3},\, \nonumber\\
&&K = 240 \pm 40~{\rm MeV}\,, \quad E_{\rm sym} = 31~{\rm MeV}\,. \label{nmp}
\end{eqnarray}
We choose the value of $m_0$ in the range of $500-900$~MeV~\cite{Motohiro:2015taa}.
Since the value of the incompressibility $K$ is relatively not well fixed compared to the other nuclear matter properties,
we use two different values of $K$ as inputs.
The determined parameters are shown in Table~\ref{paraset1}, where $K=240$~MeV, and Table~\ref{paraset2}, where $K=215$~MeV.
\begin{table}[h]
\caption{Parameter set 1: $K = 240$~MeV.}
\begin{tabular}{c | r r r r}
\hline
\hline
 ~$m_0$ (MeV)~            & ~~~$600$~~~ & ~~~$700$~~~ & ~~~$800$~~~ & ~~~$900$~~~\\
\hline
 $g_1$                           & $14.836$~        & $14.171$~        & $13.349$~          & $12.329$~\\
 $g_2$                           & $8.427$~          & $7.762$~         & $6.941$~           & $5.921$~\\
 $g_{\omega N\!N}$       & $9.132$~         & $7.305$~          & $5.660$~            & $3.522$~\\
 $g_{\rho N\!N}$            & $3.927$~         & $4.065$~          & $4.149$~           & $4.218$~\\
 $\bar{\mu}^2/f_\pi^2$ & $21.821$~        & $18.842$~        & $11.693$~         & $1.537$~\\
 $\lambda$                    & $39.367$~       & $34.584$~         & $22.578$~         & $4.388$~\\
 $\lambda_6 f_\pi^2$     & $15.344$~       & $13.540$~         & $8.683$~          & $0.649$~\\
 $m_\sigma$ (MeV)        & $411.299$~      & $385.805$~       & $330.440$~       & $269.255$~\\
\hline
\hline
\end{tabular}
\label{paraset1}
\end{table}
\begin{table}[h]
\caption{Parameter set 2: $K = 215$~MeV.}
\begin{tabular}{c| r r r r}
\hline
\hline
 ~$m_0$ (MeV)~            & ~~~$600$~~~ & ~~~$700$~~~ & ~~~$800$~~~ & ~~~$900$~~~\\
\hline
 $g_1$                           & $14.836$~        & $14.171$~      & $13.349$~         & $12.329$~\\
 $g_2$                           & $8.427$~           & $7.762$~         & $6.941$~           & $5.921$~\\
 $g_{\omega N\!N}$       & $8.902$~          & $7.055$~         & $5.471$~            & $3.389$~\\
 $g_{\rho N\!N}$            & $3.948$~           & $4.080$~         & $4.157$~           & $4.221$~\\
 $\bar{\mu}^2/f_\pi^2$ & $23.377$~         & $20.980$~        & $13.346$~         & $2.502$~\\
 $\lambda$                    & $42.369$~         & $38.921$~       & $26.128$~          & $6.673$~\\
 $\lambda_6 f_\pi^2$     & $16.790$~         & $15.739$~       & $10.580$~          & $1.969$~\\
 $m_\sigma$ (MeV)        & $413.612$~       & $384.428$~      & $324.007$~       & $257.583$~\\
\hline
\hline
\end{tabular}
\label{paraset2}
\end{table}

We remark here that the coefficient of the six-point interaction of
the $\sigma$ meson is positive, which implies that the potential is not bounded below and the system is not stable for infinite scalar mean field.
However, since the $\sigma$ field in our model is the chiral partner of the pion field, the mean value  of the $\sigma$ field in dense matter is in general smaller than the
one in free space due to (partial) chiral symmetry restoration. Therefore, our system will be stable within the mean field approximation in dense matter.
It can be seen from Tables \ref{paraset1} and \ref{paraset2} 
that the couplings of the $\sigma$ field and $\omega$ field to nucleons decrease as $m_0$ increases, which is consistent with the observation made in  Ref.~\cite{Harada:2019oaq}.

\begin{figure}[h]
\includegraphics[width=0.35\textwidth]{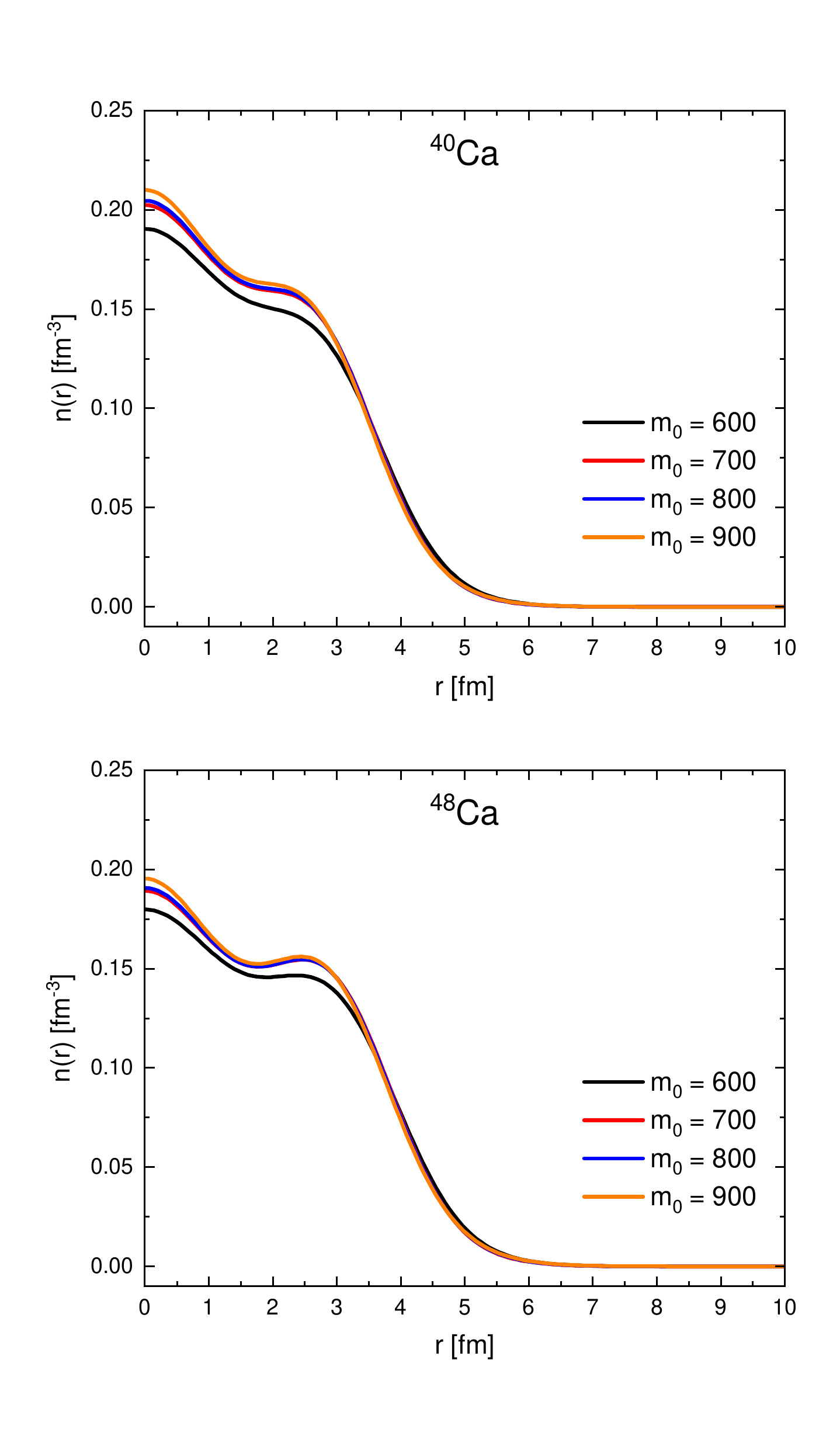}
\caption{(Color online) Nucleon density profile in $^{40}$Ca and $^{48}$Ca calculated with the parameter set 2.
}\label{fig1}
\end{figure}

In Fig.~\ref{fig1} we show nucleon density distribution $n(r)$ in $^{40}$Ca and $^{48}$Ca calculated with the parameter set 2.
From Fig.~\ref{fig1} we observe that the central density tends to increase with $m_0$.

\begin{figure}[h]
\includegraphics[width=0.4\textwidth]{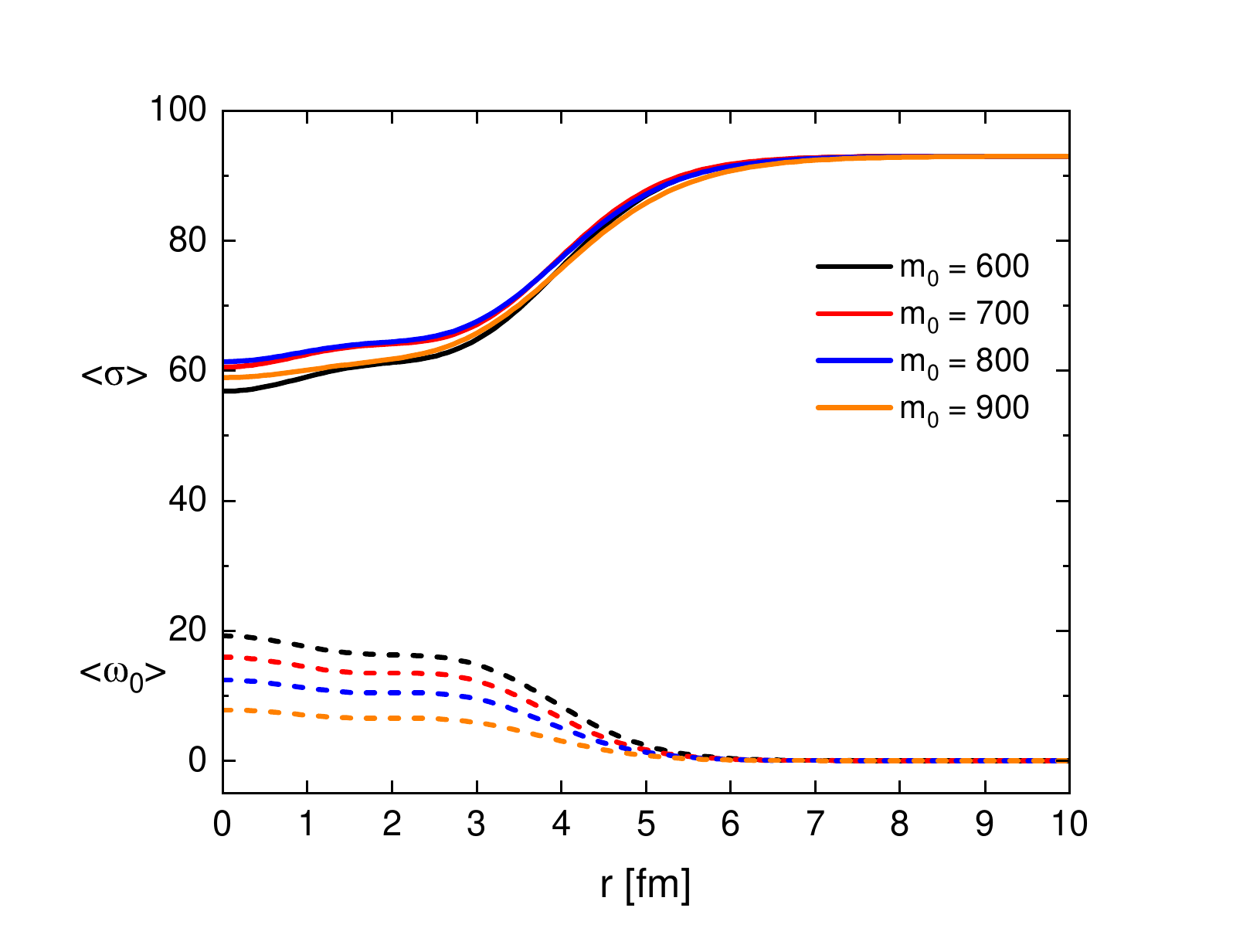}
\caption{(Color online) $\langle\sigma\rangle$ and $\langle\omega_0\rangle$  calculated with the parameter set 2 in $^{48}$Ca.
}\label{figSO}
\end{figure}

\begin{figure}[h]
\includegraphics[width=0.45\textwidth]{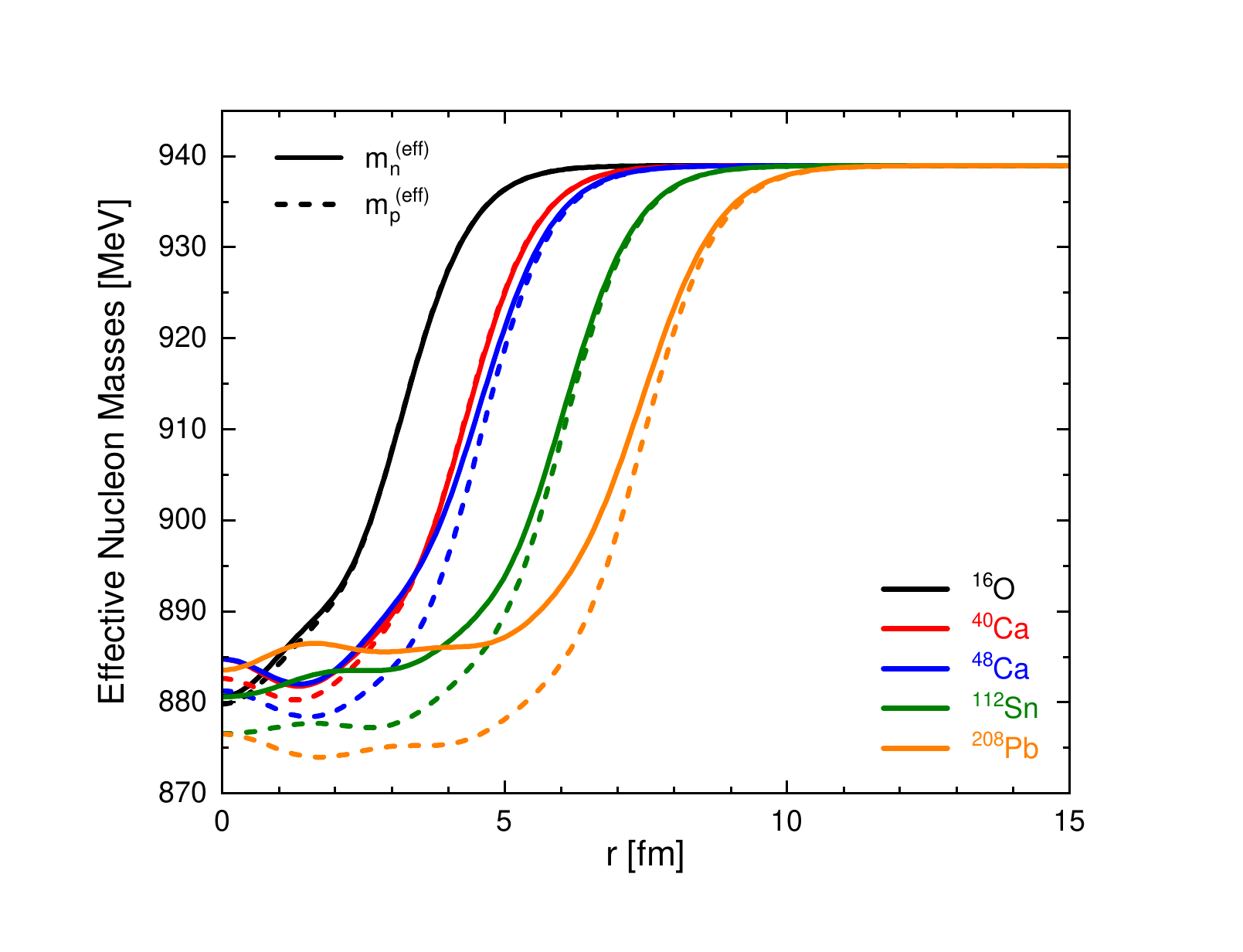}
\caption{(Color online) The neutron and proton masses in a nucleus calculated with the parameter set 3.
}\label{figNP}
\end{figure}

Now, we calculate the binding energy and the charge radius of some selected nuclei
using the parameter sets. We present our results in Tables~\ref{res1} and \ref{res2} together with the corresponding root-mean-square (RMS) deviations. 
We have used the experimental values compiled in Refs.~\cite{Wang, Xia:2017zka}.  We remark here that we cannot obtain converged numbers in our calculations
for several nuclei when $m_0=500$ MeV, and therefore we have ruled out the case with $m_0=500$ MeV.
  As it can be seen from Tables~\ref{res1} and \ref{res2}, our results approach the experimental values as $m_0$ is increased till $m_0=700$~MeV and start to deviate more afterwards.  From Tables~\ref{res1} and \ref{res2}, we conclude that $m_0=700$~MeV is favored by nuclear properties such as the nuclear binding energies and charge radii.
\begin{table}[h]
\caption{The binding energy per nucleon and the charge radius ($R_C$) with the parameter set 1.}
\begin{tabular}{c | c | c c c c | c}
\hline
\hline
~$m_0$ (MeV)~                   &                     & ~$600$~ & ~$700$~ & ~$800$~ & ~$900$~~ & ~Exp.~ \\
\hline
\multirow{5}{*}{BE/A (MeV)}  & $^{16}$O	    & $7.087$ 	& $7.280$ 	& $6.792$ 	& $5.093$ 	& $7.976$ \\
                             & $^{40}$Ca	& $7.736$ 	& $7.906$ 	& $7.538$ 	& $6.191$ 	& $8.551$ \\
                             & $^{48}$Ca   	& $7.676$ 	& $7.768$ 	& $7.378$ 	& $6.061$ 	& $8.667$ \\
                             & $^{58}$Ni	& $7.391$ 	& $7.486$ 	& $7.108$ 	& $5.849$ 	& $8.732$ \\
							 & $^{70}$Ge	& $7.761$ 	& $7.900$ 	& $7.584$ 	& $6.429$ 	& $8.722$ \\
                             & $^{82}$Se	& $7.799$ 	& $7.899$ 	& $7.580$ 	& $6.462$ 	& $8.693$ \\
                             & $^{92}$Mo	& $7.741$ 	& $7.821$ 	& $7.507$ 	& $6.424$ 	& $8.658$ \\
                             & $^{112}$Sn	& $7.668$ 	& $7.760$ 	& $7.474$ 	& $6.460$ 	& $8.514$ \\
                             & $^{126}$Sn	& $7.757$ 	& $7.801$ 	& $7.516$ 	& $6.536$ 	& $8.443$ \\
                             & $^{138}$Ba	& $7.695$ 	& $7.758$ 	& $7.482$ 	& $6.526$ 	& $8.393$ \\
                             & $^{154}$Sm	& $7.596$ 	& $7.691$ 	& $7.447$ 	& $6.540$ 	& $8.227$ \\
                             & $^{170}$Er	& $7.526$ 	& $7.587$ 	& $7.354$ 	& $6.484$ 	& $8.112$ \\
                             & $^{182}$W	& $7.418$ 	& $7.466$ 	& $7.237$ 	& $6.387$ 	& $8.018$ \\
                             & $^{202}$Pb	& $7.277$ 	& $7.303$ 	& $7.062$ 	& $6.221$ 	& $7.882$ \\
                             & $^{208}$Pb	& $7.306$ 	& $7.322$ 	& $7.075$ 	& $6.232$ 	& $7.867$ \\
\hline                             
    RMS deviation        & & $0.827$ 	& $0.737$ 	& $1.047$ 	& $2.147$ 	& $-$ \\
\hline
\multirow{5}{*}{$R_C$ (fm)} & $^{16}$O	    & $2.877$ 	& $2.792$ 	& $2.790$ 	& $2.803$ 	& $2.699$ \\
                            & $^{40}$Ca	    & $3.572$ 	& $3.491$ 	& $3.485$ 	& $3.479$ 	& $3.478$ \\
                            & $^{48}$Ca   	& $3.605$ 	& $3.537$ 	& $3.532$ 	& $3.522$ 	& $3.478$ \\
                            & $^{58}$Ni	    & $3.932$ 	& $3.863$ 	& $3.861$ 	& $3.855$ 	& $3.776$ \\
							& $^{70}$Ge	    & $4.104$ 	& $4.028$ 	& $4.018$ 	& $4.001$ 	& $4.041$ \\
                            & $^{82}$Se	    & $4.223$ 	& $4.154$ 	& $4.145$ 	& $4.125$ 	& $4.140$ \\
                            & $^{92}$Mo	    & $4.418$ 	& $4.351$ 	& $4.347$ 	& $4.335$ 	& $4.315$ \\
                            & $^{112}$Sn	& $4.684$ 	& $4.616$ 	& $4.608$ 	& $4.591$ 	& $4.594$ \\
                            & $^{126}$Sn	& $4.764$ 	& $4.703$ 	& $4.695$ 	& $4.675$ 	& $4.685$ \\
                            & $^{138}$Ba	& $4.928$ 	& $4.865$ 	& $4.856$ 	& $4.834$ 	& $4.838$ \\
                            & $^{154}$Sm	& $5.117$ 	& $5.045$ 	& $5.031$ 	& $5.004$ 	& $5.105$ \\
                            & $^{170}$Er	& $5.250$ 	& $5.181$ 	& $5.169$ 	& $5.144$ 	& $5.279$ \\
                            & $^{182}$W	    & $5.374$ 	& $5.305$ 	& $5.294$ 	& $5.270$ 	& $5.356$ \\
                            & $^{202}$Pb	& $5.555$ 	& $5.493$ 	& $5.485$ 	& $5.462$ 	& $5.471$ \\
                            & $^{208}$Pb	& $5.588$ 	& $5.529$ 	& $5.521$ 	& $5.499$ 	& $5.501$ \\
\hline                             
    RMS deviation         &  & $0.097$ 	& $0.052$ 	& $0.053$ 	& $0.062$ 	& $-$ \\
\hline
\hline
\end{tabular}
\label{res1}
\end{table}
\begin{table}[h]
\caption{The binding energy per nucleon and the charge radius ($R_C$) with the parameter set 2.}
\begin{tabular}{c | c | c c c c| c}
\hline
\hline
~$m_0$ (MeV)~                   &                      & ~$600$~ & ~$700$~ & ~$800$~ & ~$900$~~ & ~Exp.~ \\
\hline
\multirow{5}{*}{BE/A (MeV)}  & $^{16}$O	    & $7.489$ 	& $7.781$ 	& $7.298$ 	& $5.698$ 	& $7.976$ \\
                             & $^{40}$Ca	& $8.063$ 	& $8.301$ 	& $7.942$ 	& $6.693$ 	& $8.551$ \\
                             & $^{48}$Ca   	& $7.978$ 	& $8.134$ 	& $7.757$ 	& $6.541$ 	& $8.667$ \\
                             & $^{58}$Ni	& $7.685$ 	& $7.841$ 	& $7.473$ 	& $6.308$ 	& $8.732$ \\
							 & $^{70}$Ge	& $8.044$ 	& $8.239$ 	& $7.932$ 	& $6.866$ 	& $8.722$ \\
                             & $^{82}$Se	& $8.066$ 	& $8.219$ 	& $7.910$ 	& $6.881$ 	& $8.693$ \\
                             & $^{92}$Mo	& $7.993$ 	& $8.123$ 	& $7.822$ 	& $6.828$ 	& $8.658$ \\
                             & $^{112}$Sn	& $7.911$ 	& $8.050$ 	& $7.774$ 	& $6.844$ 	& $8.514$ \\
                             & $^{126}$Sn	& $7.980$ 	& $8.070$ 	& $7.802$ 	& $6.909$ 	& $8.443$ \\
                             & $^{138}$Ba	& $7.920$ 	& $8.028$ 	& $7.764$ 	& $6.890$ 	& $8.393$ \\
                             & $^{154}$Sm	& $7.821$ 	& $7.958$ 	& $7.724$ 	& $6.894$ 	& $8.227$ \\
                             & $^{170}$Er	& $7.733$ 	& $7.837$ 	& $7.618$ 	& $6.830$ 	& $8.112$ \\
                             & $^{182}$W	& $7.616$ 	& $7.707$ 	& $7.494$ 	& $6.726$ 	& $8.018$ \\
                             & $^{202}$Pb	& $7.468$ 	& $7.535$ 	& $7.310$ 	& $6.549$ 	& $7.882$ \\
                             & $^{208}$Pb	& $7.496$ 	& $7.552$ 	& $7.321$ 	& $6.558$ 	& $7.867$ \\
\hline                             
    RMS deviation        &  & $0.573$ 	& $0.438$ 	& $0.727$ 	& $1.734$ 	& $-$ \\
\hline
\multirow{5}{*}{$R_C$ (fm)} & $^{16}$O	    & $2.845$ 	& $2.763$ 	& $2.772$ 	& $2.796$ 	& $2.699$ \\
                            & $^{40}$Ca	    & $3.546$ 	& $3.469$ 	& $3.473$ 	& $3.479$ 	& $3.478$ \\
                            & $^{48}$Ca   	& $3.585$ 	& $3.521$ 	& $3.525$ 	& $3.527$ 	& $3.478$ \\
                            & $^{58}$Ni	    & $3.912$ 	& $3.848$ 	& $3.856$ 	& $3.863$ 	& $3.776$ \\
							& $^{70}$Ge	    & $4.085$ 	& $4.013$ 	& $4.013$ 	& $4.008$ 	& $4.041$ \\							
                            & $^{82}$Se	    & $4.209$ 	& $4.145$ 	& $4.144$ 	& $4.135$ 	& $4.140$ \\
                            & $^{92}$Mo	    & $4.401$ 	& $4.339$ 	& $4.344$ 	& $4.344$ 	& $4.315$ \\
                            & $^{112}$Sn	& $4.671$ 	& $4.608$ 	& $4.609$ 	& $4.602$ 	& $4.594$ \\
                            & $^{126}$Sn	& $4.754$ 	& $4.697$ 	& $4.698$ 	& $4.688$ 	& $4.685$ \\
                            & $^{138}$Ba	& $4.920$ 	& $4.862$ 	& $4.861$ 	& $4.849$ 	& $4.838$ \\
                            & $^{154}$Sm	& $5.111$ 	& $5.045$ 	& $5.039$ 	& $5.022$ 	& $5.105$ \\
                            & $^{170}$Er	& $5.242$ 	& $5.178$ 	& $5.175$ 	& $5.160$ 	& $5.279$ \\
                            & $^{182}$W	    & $5.364$ 	& $5.301$ 	& $5.298$ 	& $5.286$ 	& $5.356$ \\
                            & $^{202}$Pb	& $5.549$ 	& $5.493$ 	& $5.493$ 	& $5.481$ 	& $5.471$ \\
                            & $^{208}$Pb	& $5.584$ 	& $5.531$ 	& $5.532$ 	& $5.519$ 	& $5.501$ \\  
\hline                             
    RMS deviation         &  & $0.082$ 	& $0.046$ 	& $0.049$ 	& $0.056$ 	& $-$ \\                            
\hline
\hline
\end{tabular}
\label{res2}
\end{table}
We can understand our conclusion from the following two observations. 
First, from  Tables~\ref{paraset1} and \ref{paraset2}, 
we can see that the values of 
$\bar{\mu}^2/f_\pi^2$, $\lambda$, and $\lambda_6 f_\pi^2$ decrease slightly as $m_0$ changes from $600$ MeV to $700$ MeV,
while the values change drastically when $m_0$ is larger than $700$ MeV. Second, from Fig.~\ref{figSO} it can be seen that
$\langle\omega_0\rangle$  decreases gradually as $m_0$ increases, while $\langle\sigma\rangle$ shows a peculiar behavior.
As $m_0$ increases from $600$ MeV to $700$ MeV, $\langle\sigma\rangle$ increases, while the value changes little when 
$m_0$ increases from $700$ MeV to $800$ MeV and then drops when $m_0=900$ MeV.
From the above observation, we fix the value of the chiral invariant mass as $m_0=700$~MeV and try to improve the results of nuclear properties.
For this we use the following nuclear matter properties 
\begin{eqnarray}
&&\frac{E}{A} - m_N = -16.3~{\rm MeV}\,, \quad n_0 = 0.16~{\rm fm}^{-3},\, \nonumber\\
&&K = 215~{\rm MeV}\,, \quad\quad E_{\rm sym} = 30~{\rm MeV}\,, \label{con-paraset3}
\end{eqnarray}
and determine our model parameters again, which are given in Table~\ref{paraset3}. 
With this new parameter set, we calculate the properties of selected nuclei and compare our results with experiments in Table~\ref{res3}, where 
the RMS deviation of our results is $0.204$ for the binding energy and $0.045$ for the charge radius.
As in Table~\ref{res3}, our results are in quantitative agreement with experiments. 
To see how our results compare with the ones from a Walecka-type mean field model, 
we also show the results from, for example, 
the spherical relativistic continuum Hartree-Bogoliubov (RCHB) theory with the relativistic density functional PC-PK1~\cite{Zhao:2010hi, Xia:2017zka}.
\begin{table}[h]
\caption{Parameter set 3. All parameters are dimensionless except $m_\sigma$.}
\begin{tabular}{c c c c c c c c}
\hline
\hline
$g_1$     &$g_2$    &\:$g_{\omega NN}$ &\:$g_{\rho NN}$ &\:$\bar{\mu}^2/f_\pi^2$ &$\lambda$ &\:$\lambda_6 f_\pi^2$ &$m_\sigma$ [MeV]\\
\hline
$14.171$ &$7.762$ &\:$7.036$               &\:$3.958$           &\:$21.135$                      &$39.332$   &\:$15.996$                  &\:$382.140$\\
\hline
\hline
\end{tabular}
\label{paraset3}
\end{table}
\begin{table}[h]
\caption{The binding energy per nucleon and the charge radius ($R_C$) with the parameter set 3.}
\begin{tabular}{c | c  c  c | c  c  c }
\hline
\hline
\multirow{2}{*}{} &\multicolumn{3}{c|}{BE/A (MeV)}     & \multicolumn{3}{c}{$R_C$ (fm)} \\
\cline{2-7}
              & PDM       & RCHB      & Exp.      & PDM       & RCHB      & Exp. \\
\hline
 $^{16}$O	    & $8.040$    & $7.956$    & $7.976$    & $2.757$    & $2.768$    & $2.699$ \\
 $^{40}$Ca	    & $8.574$    & $8.577$    & $8.551$    & $3.464$    & $3.481$    & $3.478$ \\
 $^{48}$Ca      & $8.419$    & $8.654$    & $8.667$    & $3.517$    & $3.494$    & $3.478$ \\
 $^{58}$Ni	    & $8.118$    & $8.691$    & $8.732$    & $3.843$    & $3.737$    & $3.776$ \\
 $^{70}$Ge	    & $8.521$    & $8.650$    & $8.722$    & $4.010$    & $4.001$    & $4.041$ \\
 $^{82}$Se	    & $8.513$    & $8.664$    & $8.693$    & $4.142$    & $4.125$    & $4.140$ \\
 $^{92}$Mo	    & $8.408$    & $8.662$    & $8.658$    & $4.335$    & $4.310$    & $4.315$ \\
 $^{112}$Sn     & $8.339$    & $8.489$    & $8.514$    & $4.605$    & $4.582$    & $4.594$ \\
 $^{126}$Sn     & $8.372$    & $8.447$    & $8.443$    & $4.695$    & $4.683$    & $4.685$ \\
 $^{138}$Ba     & $8.329$    & $8.406$    & $8.393$    & $4.860$    & $4.848$    & $4.838$ \\
 $^{154}$Sm     & $8.263$    & $8.149$    & $8.227$    & $5.042$    & $5.062$    & $5.105$ \\
 $^{170}$Er     & $8.140$    & $8.000$    & $8.112$    & $5.176$    & $5.224$    & $5.279$ \\
 $^{182}$W	    & $8.007$    & $7.927$    & $8.018$    & $5.299$    & $5.342$    & $5.356$ \\
 $^{202}$Pb     & $7.837$    & $7.869$    & $7.882$    & $5.491$    & $5.490$    & $5.471$ \\
 $^{208}$Pb     & $7.860$    & $7.875$    & $7.867$    & $5.529$    & $5.518$    & $5.501$ \\
 \hline                             
    RMS deviation         & $0.204$ & $0.05$  	& $-$	& $0.045$ 	& $0.031$ 	& $-$ \\
\hline
\hline
\end{tabular}
\label{res3}
\end{table}
Using the parameter set 3, we finally calculate the $r$-dependence of 
neutron and proton masses 
 which are given by Eq.~(\ref{mass}) with the $r$-dependent $\langle \sigma \rangle$, 
and present our results in
Fig.~\ref{figNP}. Our results show that, as expected, the neutron-proton mass difference is larger in the nuclear system with a
larger isospin asymmetry.
 
\section{Summary}\label{summary}
In this work, using the extended parity doublet model~\cite{Motohiro:2015taa}, we calculated the properties of some stable nuclei in the mean field approximation to estimate
the value of the chiral invariant mass preferred by nuclear properties.
 Since our primary goal in this work is to see if the parity doublet model can explain some basic nuclear properties such as the binding energy, we didn't consider pairing correlations which are essential for odd-even staggering in nuclear properties.
We observed that our results are closest to the experiments when we take $m_0=700$~MeV.
Therefore, we have concluded that the chiral invariant mass contribution to
the nucleon mass is around $700$~MeV, which can be understood from the peculiar behavior of $\bar{\mu}^2/f_\pi^2$, $\lambda$, $\lambda_6 f_\pi^2$, $\langle\sigma\rangle$, and $\langle\omega_0\rangle$ as the value of $m_0$ changes.
We also calculated the neutron and proton masses in a nucleus and observed that the neutron-proton mass difference
becomes larger in a isospin asymmetric nucleus.

In the future we will extend our present study by including pairing correlations and deformations
 using more complete approaches such as (deformed) RCHB theory which
can provide an appropriate treatment of the pairing correlation in the presence of the continuum through the Bogoliubov transformation in a microscopic and self-consistent way~\cite{Meng:2015hta}.
\section*{Acknowledgments}
We thank Jie Meng for providing the RCHB code.
This work was supported in part by the Rare Isotope Science Project of Institute for Basic Science funded by Ministry of Science and ICT and National Research Foundation of Korea (2013M7A1A1075764), by the Institute for Basic Science (IBS-R031-D1), by National Research Foundation of Korea
(NRF) grants funded by the Korean government (Ministry of Science and ICT) (No. 2021R1F1A1060066, No.2020R1A2C3006177), and by JPSP KAKENHI Grant Number 
 20K03927.


\begin{thebibliography}{99}
\bibitem{Detar:1988kn}
  C.~E.~Detar and T.~Kunihiro,
  Phys.\ Rev.\ D {\bf 39}, 2805 (1989).

\bibitem{Hatsuda:1988mv}
  T.~Hatsuda and M.~Prakash,
  Phys.\ Lett.\ B {\bf 224}, 11 (1989).

\bibitem{Zschiesche:2006zj}
  D.~Zschiesche, L.~Tolos, J.~Schaffner-Bielich and R.~D.~Pisarski,
  Phys.\ Rev.\ C {\bf 75}, 055202 (2007).

\bibitem{Dexheimer:2007tn}
  V.~Dexheimer, S.~Schramm and D.~Zschiesche,
  Phys.\ Rev.\ C {\bf 77}, 025803 (2008).

\bibitem{Sasaki:2010bp}
  C.~Sasaki and I.~Mishustin,
  Phys.\ Rev.\ C {\bf 82}, 035204 (2010).

\bibitem{Gallas:2011qp}
  S.~Gallas, F.~Giacosa and G.~Pagliara,
  Nucl.\ Phys.\ A {\bf 872}, 13 (2011).

\bibitem{Sasaki:2011ff}
C.~Sasaki, H.~K.~Lee, W.~G.~Paeng and M.~Rho,
Phys. Rev. D \textbf{84}, 034011 (2011).

\bibitem{Steinheimer:2011ea}
  J.~Steinheimer, S.~Schramm and H.~Stocker,
  Phys.\ Rev.\ C {\bf 84}, 045208 (2011).

\bibitem{Benic:2015pia}
  S.~Benic, I.~Mishustin and C.~Sasaki,
  Phys.\ Rev.\ D {\bf 91}, no. 12, 125034 (2015).

\bibitem{Motohiro:2015taa}
  Y.~Motohiro, Y.~Kim and M.~Harada,
  Phys.\ Rev.\ C {\bf 92}, no. 2, 025201 (2015);
  Erratum: [Phys.\ Rev.\ C {\bf 95}, no. 5, 059903 (2017)]

%
\bibitem{Harada:2019oaq}
M.~Harada and T.~Yamazaki, ``Charmed Mesons in Nuclear Matter Based on Chiral Effective Models,''
JPS Conf. Proc. \textbf{26}, 024001 (2019).

\bibitem{Bender:2003jk}
  M.~Bender, P.~H.~Heenen and P.~G.~Reinhard,
  Rev.\ Mod.\ Phys.\  {\bf 75}, 121 (2003).

\bibitem{Vretenar:2005zz}
  D.~Vretenar, A.~V.~Afanasjev, G.~A.~Lalazissis and P.~Ring,
  Phys.\ Rept.\  {\bf 409}, 101 (2005).

\bibitem{Meng:2005jv}
  J.~Meng, H.~Toki, S.~G.~Zhou, S.~Q.~Zhang, W.~H.~Long and L.~S.~Geng,
  Prog.\ Part.\ Nucl.\ Phys.\  {\bf 57}, 470 (2006).

\bibitem{Meng:2016}
  {\it Relativistic Density Functional for Nuclear Structure},
  International Review of Nuclear Physics, Vol.~10,
  edited by J.~Meng (World Scientific, Singapore, 2016).

\bibitem{Schunck:2019}
  {\it Energy Density Functional Methods for Atomic Nuclei},
  edited by N. Schunck (IOP Publishing Ltd 2019).

\bibitem{Jido:2001nt}
  D.~Jido, M.~Oka and A.~Hosaka,
  Prog.\ Theor.\ Phys.\  {\bf 106}, 873 (2001).

\bibitem{Gallas:2009qp}
  S.~Gallas, F.~Giacosa and D.~H.~Rischke,
  Phys.\ Rev.\ D {\bf 82}, 014004 (2010).

\bibitem{Paeng:2011hy}
  W.~G.~Paeng, H.~K.~Lee, M.~Rho and C.~Sasaki,
  Phys.\ Rev.\ D {\bf 85}, 054022 (2012).

\bibitem{Bando:1987br}
  M.~Bando, T.~Kugo and K.~Yamawaki,
  Phys.\ Rept.\  {\bf 164}, 217 (1988).

\bibitem{Harada:2003jx}
  M.~Harada and K.~Yamawaki,
  Phys.\ Rept.\  {\bf 381}, 1 (2003).

\bibitem{Wang}
  M.~Wang {\it et al.},
  Chin.\ Phys.\ C {\bf 36}, 1603 (2012).

\bibitem{Xia:2017zka}
  X.~W.~Xia {\it et al.},
  Atom.\ Data Nucl.\ Data Tabl.\ {\bf 121-122}, 1 (2018).

  \bibitem{Zhao:2010hi}
  P.~W.~Zhao, Z.~P.~Li, J.~M.~Yao and J.~Meng,
  Phys.\ Rev.\ C {\bf 82}, 054319 (2010).
  

\bibitem{Meng:2015hta}
  J.~Meng and S.~G.~Zhou,
  J.\ Phys.\ G {\bf 42}, no. 9, 093101 (2015).

\end{thebibliography}
\end{document}